\documentstyle[12pt,epsf]{article}
\begin{document}
\baselineskip=4.5mm
\pagestyle{empty}

\centerline{\Large\bf Classical and Quantum Chiral Order}
\centerline{\Large\bf in Frustrated {\it XY\/} Magnets}

\bigskip
\centerline{Hikaru Kawamura}

\bigskip
\centerline{\sl Department of Earth and Space Science, Faculty of Science,}
\centerline{\sl Osaka University, Toyonaka 560-0043, Japan}


\begin{abstract}
\textwidth 4.2in
\baselineskip=4.5mm
Recent studies on the chiral order of regularly frustrated {\it XY\/}
magnets are
reviewed both in classical and quantum  cases.
In the classical case, chiral transition is a thermal one,
while in the quantum case, it is a quantum phase transition.
Importance of spatial dimensionality  on the chiral order is
clarified. Particular attention is paid to the
possible ``spin-chirality decoupling'' phenomenon,
and to the possible  pure chiral phase of either thermal
or quantum origin where the chirality exhibits a long-range order
without the standard spin order.
\end{abstract}


\textwidth 6.2in
\bigskip
\noindent
{\bf 1. Introduction}
\medskip

Magnetic ordering of geometrically frustrated antiferromagnets
has attracted continual interest of researchers in magnetism
and statistical physics\cite{Diep,Collins,Kawareview,Ramirezreview,
Schifferreview,Harrisreview,Kawareview2}.
In geometrically frustrated
antiferromagnets,  spins usually
sit on lattices made up of triangles
as elementary units, and interact antiferromagnetically
with their neighboring spins. Intrinsic inability to simultaneously satisfy all
antiferromagnetic nearest-neighbor interactions on a triangle
necessarily leads to macroscopic frustration. This makes 
the spin ordering on these lattices a highly nontrivial issue.
Recent studies have revealed that
frustration often gives rise to new interesting phenomena
in the magnetic ordering,
{\it e.g.\/},  phase transitions of novel universality classes, exotic
ordered phases with novel order parameter, and the spin-liquid phase stabilized
down to extremely low temperatures, {\it etc\/}.
In this short review, I wish to deal with both the {\it classical\/}
and {\it quantum\/} ``chiral'' phase transitions realized in certain frustrated
{\it XY\/}-like antiferromagnets.

One interesting consequence of spin
frustration in vector spin systems is the possible
appearance of ``chiral'' degrees of freedom
\cite{Kawareview,Kawareview2,Villain}.
``Chirality'' is a multispin quantity representing the sense or the
handedness of the noncollinear spin structures induced by
spin frustration.
Two different types of chiralities have  often been discussed
in the literature: One is a {\it vector chirality\/} and the other
a {\it scalar chirality\/}.

Chiral states representing the right- and left-handed configurations
are illustrated in Fig.1 for an example of three
antiferromagnetically coupled {\it XY\/} spins
located at each corner of a triangle. The ground-state
spin configuration
is a well-known $120^\circ$ spin structure, in which each
{\it XY\/} spin on a plane
makes an angle equal to $\pm 120^\circ$ with the neighboring
spins. Here, one may define
the chirality of the first type, the vector chirality,
via a vector product of the two neighboring
spins, averaged over three spin pairs, by
\begin{equation}
\kappa =\frac{2}{3\sqrt 3}\sum _{<ij>}
\left [ \vec S_i\times \vec S_j\right ]_z,
\end{equation}
where the summation is taken over three pairs of sites along the sides
of the triangle in a clockwise direction.
Evidently, the sign of $\kappa $ represents each of the two
chiral states,
{\it i.e.\/}, either a
right-handed (clockwise) state for $\kappa >0$ or a
left-handed (counterclockwise) state for $\kappa <0$.
In the case of {\it XY\/} spins, the vector chirality
$\kappa $ is actually a
{\it pseudoscalar\/} from a symmetry viewpoint:
It remains invariant under global
$SO(2)=U(1)$ proper spin rotations while it changes sign under
global $Z_2$ spin reflections. Hence,
in order to transform one chiral state
to the other, one needs to make a mirroring operation, {\it i.e.\/},
a global spin reflection. The
chiral order is then closely related
to the spontaneous breaking of a discrete $Z_2$ spin-reflection
symmetry.

\begin{figure}[ht]
\begin{center}
\noindent
\epsfxsize=0.63\textwidth
\epsfbox{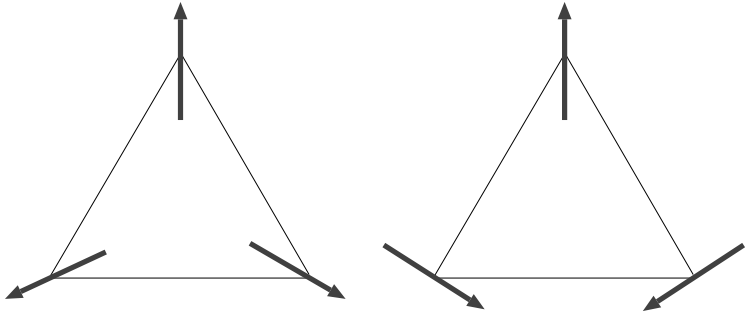}
\end{center}
\caption{Two chiral states in the ground-state spin
configurations of antiferromagnetically coupled three {\it XY\/} spins
on a triangle. These two chiral states are characterized by the
mutually opposite signs of the vector chirality.
}
\end{figure}

In the case of
three-component Heisenberg spins, the second
type of chirality, the scalar chirality, has  been
discussed. It is defined
for three neighboring Heisenberg spins by
$\chi =\vec S_1\cdot \vec S_{2} \times \vec S_{3}$.
This scalar chirality takes a nonzero value for noncoplanar spin
states, its sign representing whether the noncoplanar structure is
either right- or left-handed. Note that, in contrast to the
vector chirality,  the scalar chirality vanishes
for any coplanar spin structure even when it is noncollinear.
This scalar chirality is invoked in recent studies of
magnetic properties of geometrically frustrated antiferromagnets
\cite{KawaAri}
and spin glasses\cite{Kawareview2,KawaSG1,KawaSG2,HukuKawa,
KawaImaSG}, but also of
transposrt properties 
in manganites or pyrochlore magnets\cite{Ye,Chun,Nagaosa,
Taguchi,Sato}.

Our main concern in this article is the ordering
of geometrically frustrated {\it XY\/} antiferromagnets.
Hence,  we consider
in the following possible chiral order {\it associated with the
vector chirality\/}. We also focus in this article
on the chiral order in
{\it regularly\/} frustrated systems without any quenched randomness.
Historically, possible
chiral order of frustrated {\it XY\/} antiferromagnets
have been studied mainly for  classical systems
as a thermal
ordering phenomenon. Thus, we first review in \S 2
the approaches performed
for the antiferromagnetic {\it XY\/} models on various triangle-based
lattices in one, two and three spatial dimensions, {\it i.e.\/},
the one-dimensional (1D) triangular-ladder lattice, the two-dimensional (2D)
triangular lattice and the three-dimensional (3D) stacked-triangular
lattice. All these lattices consist of triangles as elementary units,
and the
antiferromagnetic {\it XY\/} models defined on these lattices
possess nontrivial chiral degrees of freedom as illustrated in Fig.1.
As is well-known in theory of critical phenomena,
the spatial dimensionality is crucially
important in determining the nature of phase transition.
Indeed, it has turned out that
chiral order largely changes its nature
depending on the spatial dimensionality of the lattice.

In the last decade,
quantum phase transitions and quantum critical phenomena
have attracted a lot of attention in various branches
of condensed-matter and statistical
physics. Under such circumstances,
it would be quite natural to ask what is the nature of the
{\it quantum\/} chiral order, possibly realized in the ground state
of purely quantum systems.
In other words, it is possible to realize chiral
order via a pure quantum phase transition with varying
some parameters of the Hamiltonian at zero temperature?
If yes, what is its nature in comparison with the thermal chiral
order? Indeed, such studies of quantum chiral order
was made extensively in these last few years. As an example of such recent
studies,
we wish to review in \S 3 theoretical studies
on the chiral order of frustrated quantum spin chains.
Finally in \S 4, we give brief summary and discussion,
and conclude the review.
\bigskip

\noindent
{\bf 2. Chiral order in classical {\it XY\/} systems}
\medskip

In this section, we consider the thermal chiral order in purely
classical systems. In order to clarify the important role of spatial
dimensionality, we deal with the 1D, 2D and 3D triangular-lattice {\it XY\/}
models in the following subsections \S 2.1-3, respectively.

\medskip
\noindent
{\bf 2.1 One-dimensional triangular-ladder lattice}
\medskip

Let us begin with the 1D example. The model we consider is the
classical two-component {\it XY\/} (plane rotator) model on the
triangular ladder, a linear array of alternating upward and downward
triangles. Each site
has four nearest neighbors. The Hamiltonian is given by
\begin{equation}
{\cal H}=J\sum _{<ij>}\vec S_i\cdot \vec S_j,
\end{equation}
where the interaction is assumed to be antiferromagnetic ($J>0$) and
work only between nearest-neighboring spins, while $\vec S_i=(S_i^x,S_i^y)$ is
a two-component unit vector located at the $i$-th site
on the triangular ladder.
One may define the chirality at each upward triangle
on the ladder by Eq.(1).

This 1D model can be solved exactly at
arbitrary temperature by the standard technique, and
the solution was reported by Horiguchi and Morita\cite{Horiguchi}.
The ground state of this model is the $120^\circ$
spin structure with either right-handed ($\kappa=1$)
or left-handed ($\kappa=-1$) chirality.
Hence, at $T=0$, the model exhibits a full long-range order (LRO)
both in the spin and in the chiral sectors.
Meawhile, since the model is a 1D one
with short-range interaction,
both the spin-spin and the chirality-chirality correlation
functions remain short-ranged at any finite temperature without
a finite-temperature transition of any type.
Hence, the present model exhibits
a zero-temperature phase transition both in the spin and in the
chiral sectors.

The nontrivial issue is the manner how the spin and the chirality order
at $T=0$. Horiguchi and Morita
observed by exact calculations that the
spin-correlation length defined via the spin-spin correlation
function $C_s(x)=<\vec S_0\cdot \vec S_x>\approx A\exp (-x/\xi_s)$ ($A$ being
some constant) diverges with decreasing  temperature as a power law, 
characterized by the associated
spin-correlation-length exponent equal to unity, $\nu _s=1$,
\begin{equation}
\xi_s\approx T^{-1}.
\end{equation}
This divergence is common with the one observed in the unfrustrated
1D {\it XY} model.
Meanwhile, the chiral-correlation length defined via the
chirality-chirality correlation
function $C_\kappa(x)=<\kappa_0 \kappa_x>\approx A'\exp (-x/\xi_\kappa)$
($A'$ being some constant)
was found to diverge exponentially with temperature as
\begin{equation}
\xi_\kappa\approx \exp(-J/T),
\end{equation}
which means the chiral-correlation-length exponent  equal to
infinity, $\nu _\kappa =\infty$. Such an exponential
divergence happens to be common with
the one observed in the standard 1D Ising model.
A remarkable observation here is, though not necessarily be emphasized in
Ref.\cite{Horiguchi},
that the manners how the spin and the chirality correlations
grow toward the $T=0$ transition are mutually different, each characterized
by mutually distinct correlation-length exponents, $\nu _s=1$
{\it vs\/.} $\nu_\kappa=\infty$. This means that there exist two distinct
diverging length scales in this $T=0$ transition, one associated with
the {\it XY\/} spin and the other associated with the chirality.
The situation
in in sharp contrast to that of the
standard continuous (second-order) phase transitions
characterized by only one diverging length scale (one-length-scaling
hypothesis). Although the chirality is written as a product of two
{\it XY\/} spins on short
length scales of order lattice spacing,
the chirality eventually
outgrows the spin on long length scale, at least in an immediate vicinity
of the $T=0$ transition point, since $\nu_\kappa > \nu_s$ means
$\xi_\kappa >>\xi_s$.
We note that such an unusual situation, {\it i.e.\/},
the spin and the chirality exhibiting qualitatively different transition
behaviors on long length scales, entails the ``spin-chirality
decoupling'' or the ``spin-chirality separation'' on long length scale.
Although both the spin and the chirality order simulataneously
at $T=0$ reflecting the 1D character
of the model, the occurrence of the spin-chirality decoupling
leads to an apparent violation of the one-length-scaling
hypothesis, which, in turn,
enables the spin and the chirality to exhibit mutually different transition
behaviors.

In 2D, both the standard (unfrustrated)
{\it XY\/} model
and the standard Ising model
are known
to exhibit a finite-temperature transition. It would then be interesting
to see whether the spin-chirality decoupling occurs in the
frustrated {\it XY\/} model in 2D, and
if it occurs, how the both order with decreasing
temperature. Indeed, this problem has been studied quite extensively
in the past fifteen years, which we will now review in the next subsection.

\medskip
\noindent
{\bf 2.2 Two-dimensional triangular lattice}
\medskip

Typical example of the 2D frustrated {\it XY\/} model is
the antiferromagnetic {\it XY\/} model on the triangular lattice
\cite{MiyaShiba,DHLee,Himbergen,JLee,Southern,Capriotti,SLee2}.
We note that essentially similar
physics is also expected to occur in some other models such as the
fully-frustrated {\it XY\/} model on the square lattice
\cite{JLee,Teitel,Jose,SLee1,Olsson,Loison}, or its dual
counterpart (the Coulomb gas)\cite{Thijssen,Grest,JRLee}, {\it etc\/}.
While we present our discussion here in terms of
the triangular-lattice {\it XY\/} antiferromagnet,
the reader will find  in cited references
essentially similar results and controversy
for these other models as well. The field-theoretical RG analysis was 
also applied to this 2D problem\cite{Parruccini}.

The spin and chirality ordering in the antiferromagnetic triangular-lattice
{\it XY\/} model was first studied by means of Monte Carlo (MC) simulations
by Miyashita and Shiba\cite{MiyaShiba},
and by Lee, Joannopoulos and Landau\cite{DHLee}.
Miyashita and Shiba suggested that the spin and the chirality ordered
at two close but separate finite temperatures.
With decreasing temperature, the chirality ordered first at $T=T_\kappa$
characterized by the onset of the chiral LRO
keeping the spin
paramagnetic, and then at a slightly lower temperature $T=T_s<T_\kappa$,
the spin exhibited a Kosterlitz-Thouless(KT) transition below which
the quasi-LRO of {\it XY\/} spins developed and coexisted with the
chiral LRO already established at a higher temperature $T=T_\kappa$.
According to their scenario, the model exhibits a pure chiral phase
at an intermediate temperature range $T_s<T<T_\kappa$
where the chirality exhibits a true
LRO not accompanying the standard spin order.
Obviously, such an ordering behavior
requires the spin-chirality decoupling
because the spin and the chirality  order at different temperatures.
According to Miyashita and Shiba,
the criticality at the upper chiral transition at $T=T_\kappa $ was that
of the standard 2D Ising model,  with the associated chiral exponents
$\alpha =0$(log), $\beta _\kappa=1/8$, 
{\it etc\/}. Likewise, the criticality at the lower spin
transition at $T=T_s$ was found to be
of the standard KT universality,  with the estimated
spin-anomalous-dimension exponent
$\eta =0.25$ in agreement with the standard KT value.

In contrast, Lee {\it et al\/} suggested a somewhat
different scenario for the same model\cite{DHLee}.
According to these authors, the spin and the chirality ordered
at a common finite temperature $T=T_c(=T_s=T_\kappa$)
where both the chiral LRO and the spin qiasi-LRO set in simultaneously.
Yet, Lee {\it et al\/}
suggested qualitatively different divergent behaviors to occur at
$T=T_c$ for the
spin-correlation length and for the chiral-correlation-length,
the former exhibiting a power-law divergence with
an exponent $\nu_\kappa\sim 1$ and the latter exhibiting the
KT-like exponential divergence.
This means that, in spite of the simultaneous occurrence
of the spin and the chirality transitions, the model still exhibits
the spin-chirality decoupling in the sense that
there exist two distinct deverging length scales
at the transition.

Numerous numerical works  have been performed since then on the
same and related models with the aim at
clarifying the nature of the transition.
While the controversy have continued,
and this controversy has not yet been settled completely,
recent numerical works tend to converge
in that the spin and the chirality
order at two close but separate temperatures,
the chiral ordering preceding the spin
ordering, $T_\kappa>T_s$
\cite{Southern,Capriotti,SLee2,SLee1,Olsson,Loison,Grest,JRLee}. 
The estimated $T_\kappa$ and $T_s$ are mutually
close, the difference being
only of order $(T_\kappa-T_s)/T_\kappa= 0.3\%\sim 3\%$, 
depending on the particular model and the authors.
As such, it appears that  frustrated
2D {\it XY\/} models generically possess a pure chiral phase
in a narrow but finite temperature range.
(It might also be worth mentioning that
there exists a certain 2D model, a coupled Ising-{\it XY\/}-Heisenberg
model invented to mimic the superfluidity
transition of helium-three film, where the $Z_2$ and $U(1)$ orderings were
observed to occur
at widely separate temperatures, more than 8\% apart
\cite{helium}.)

The issue of criticality of the spin and of the chirality, by contrast,
remains more ambiguous.
Some authors claim that the
criticalities of the chirality and of the spin are of
the standard Ising
and KT ones\cite{Southern,Capriotti,Olsson}, respectively, 
while others claim that they
are distinct from the standard Ising and KT ones even when both order at
two distinct temperatures\cite{SLee2,SLee1,Loison,JRLee}.

\medskip
\noindent
{\bf 2.3  Three-dimensional stacked-triangular lattice}
\medskip

In this subsection, we wish to briefly touch upon the spin and the chirality
orderings of the 3D triangular-lattice
{\it XY\/} antiferromagnet.
There are several ways to construct a 3D lattice by
stacking the 2D triangular layers.
We consider here the simplest construction which preserves the
chiral $Z_2\times U(1)$ symmetry, {\it i.e.\/}, the 3D
stacked-triangular lattice (or a simple-hexagonal lattice)
in which the 2D triangular layers are
stacked in register on top of each other.
Since there is no frustration along the orthogonal direction in this
type of stacked-triangular lattice
irrespective of the sign of the interplane interaction,
the ordered-state spin configuration
is a three-sublattice  $120^\circ$ spin structure in each
triangular layer.

In 3D,  we have several experimental
realizations of the model at issue. Indeed, various
stacked-triangular antiferromagnets with nontrivial chiral
degree of freedom have been known: Examples are
ABX$_3$-type compounds CsMnBr$_3$ and CsVBr$_3$. Even Ising-like
ABX$_3$-type compounds with an easy-axis-type anisotropy,
such as CsNiCl$_3$, CsNiBr$_3$ and CsMnI$_3$,
exhibit the chiral critical behavior under external fields higher than
a certain critical field. Extensive experimental measurements have been
performed on these chiral {\it XY\/}-like antiferromagnets which have been
summarized in several review articles\cite{Diep,Collins,Kawareview,
Kawareview2}. Recent
experimental progress has  made it
possible even to directly observe the chirality
by using the polarized neutron-scattering technique\cite{Maleyev,Plakhty}.

In sharp contrast to the 1D and 2D cases,
there are good numerical and experimental
evidence in 3D that the spin and the chirality ordered
simultaneously in 3D via a single phase transition accompanied with
the onset of the noncollinear spin LRO (120$^\circ$ structure).
In particular, Plakhty observed by means of
polarized neutron-scattering
measurements on the triangular-lattice
{\it XY\/} antiferromagnet CsMnBr$_3$
that the spin and the chirality indeed ordered simultaneously\cite{Plakhty}.
The next question would then be
whether the simultaneous spin and chirality transition
accompanies the spin-chirality decoupling or not,
namely, whether $\nu_\kappa =\nu_s$ or $\nu_\kappa \neq\nu_s$
at the transition. In fact,
the values of $\nu_s$ and $\nu_\kappa$,  estimated
either numerically\cite{HKtri1}or experimentally\cite{Plakhty},
turned out to be
close to each other, 
suggesting that the equality
$\nu_\kappa =\nu_s$ is likely to hold.
Hence, in the case of the 3D stacked-triangular {\it XY\/} angiferromagnet,
there occurs a single phase transition with a common spin- and
chirality-correlation-length exponent
$\nu=\nu_\kappa=\nu_s$  without the spin-chirality decoupling.
The situation here is in sharp contrast to
the 1D and 2D cases  where the spin-chirality decoupling
takes place.
Such a difference may be understandable if one notes the following:
Since the spin-chirality decoupling does not arise in the
mean-field limit corresponding to an infinite dimension,
higher dimensionality generally tends to suppress the spin-chirality
decoupling and to recover
the conventional transition behavior with a common
diverging length scale
occurring in both the spin and in the chiral sectors.
This suggests that
strong fluctuations borne by the combined effects of low
dimensionality and frustration  should be crucial
in realizing the spin-chirality decoupling.

Even if the 3D chiral transition is conventional
in the above sense,
we note that the associated criticality
may well be non-standard.
Rather, the chiral $Z_2\times U(1)$ symmetry simultaneously
broken at the transion
might give rise to the non-standard criticality,
possibly described by a new type of fixed point (chiral fixed point).
Indeed, this was the proposal
made some time ago by the present author: On the basis of a
symmetry argument \cite{HKtri1,HKtri3}, Monte Carlo simulations
\cite{HKtri1,HKtri4} and  renormalization-group (RG) calculations\cite{HKtri2},
possible occurrence of such
new chiral universality class
was suggested for the 3D chiral {\it XY\/} system.
While various experiments\cite{Collins,Kawareview,Kawareview2,
Ajiro,Mason,Gaulin,Wang,Deutsch,Beckman,
Enderle1,Weber,Enderle2}
performed on the 3D stacked-triangular {\it XY\/} antiferromagnet
without lattice distortion ({\it e.g.\/}, CsMnBr$_3$)
generally support this conjecture,
several theoretical works claimed that the transition should
in fact be weakly
first order, and the situation remains controversial.
We donot intend here to enter into further details of the controversy,
nor to give a complete list of references.
The reader is invited to
several recent review articles.\cite{Diep,Kawareview,Kawareview2}
(Some of the very recent theoretical works have not been 
included in these reviews: Mentioning only a few of them,
six-loop RG calculation
favors the chiral-universality scenario\cite{Pelisetto}, while
the so-called ``exact RG'' calculation
favors the weak first-order transition\cite{Mouhana}, 
{\it etc\/}.)

\medskip
I wish to conclude this section by
summarizing the ordering properties of the classical chiral {\it XY\/}
systems presented in each subsection. In 1D and 2D,
the spin-chirality decoupling takes place. The spin and the chirality
show mutually different transition behaviors. This is
in contrast to the
3D case where the spin-chirality decoupling does not occur.
In 1D, both the spin and the chirality order simultaneously
at $T=0$, but with
mutually different correlation-length exponents, $\nu_\kappa>\nu_s$.
In 3D, both the spin and the chirality
order simultaneously at a finite temperature.
Unlike the 1D case,  there is only one diverging length scale
at this transition,
the spin and the chirality possessing a common correlation-length
exponent  $\nu_\kappa=\nu_s$.
In 2D, recent works strongly suggest that the spin
and the chirality order at two close but separate temperatures,
$T_\kappa>T_s$. This means that in 2D there occurs a pure chiral phase
at an intermediate temperature regime, $T_s<T<T_\kappa$, where
only the chirality exhibits a LRO keeping
the {\it XY\/}-spin paramagnetic.

\bigskip

\noindent
{\bf 3. Chiral order in quantum {\it XY\/} systems}
\medskip

It sometimes happens that the $D$-dimensional quantum system at
zero temperature can be mapped onto the $D+1$-dimensional classical
system at finite temperature. For example, thermodynamic properties
of certain $D+1$-dimensional classical
system at finite temperature embodied in the maximum eigenvalue
of the associated transfer matrix can often
be mapped onto the ground-state
properties of  appropriate $D$-dimensional quantum system.
Of course, in order to substantiate the correspondence,
such an analogy has to be examined carefully in each particular
case.
Nevertheless, one may make a first guess that various thermal
chiral order
identified in  classical systems might have some counterparts in the
corresponding quantum systems whose spatial
dimension is one-dimension less than the
classical ones.
In view of the property of the 2D classical {\it XY\/}
system reviewed in \S 2.2,
one may then
imagine that the 1D frustrated quantum {\it XY\/} spin
chain might exhibit  quantum chiral
order at $T=0$ with varying a suitably defined parameter of the
Hamiltonian.
Motivated by such an expectation, we recently undertook a
systematic
numerical investigation of the frustrated quantum {\it XY\/} spin chain
based on the exact-diagonalization and the
density-matrix-renormalization-group
(DMRG) methods\cite{KKH,HKK1,HKK2,HKKreview}.
We have then found that
the above naive expectation based on the classical-quantum analogy
does indeed hold.

Since the details of the calculations have already been given in
Refs.\cite{KKH,HKK1,HKK2} and
in a recent review article\cite{HKKreview},
we sketch here only the gross features of the
ordering properties of the model. The Hamiltonian considered is
\begin{equation}
{\cal H} = \sum_{\rho=1,2} J_\rho \sum_l
\left( S^x_l S^x_{l+\rho} + S^y_l S^y_{l+\rho} \right),
\end{equation}
where $J_1 > 0$ and $J_2 > 0$ are the antiferromagnetic nearest-neighbor
and next-nearest-neighbor couplings along the 1D chain, while $\vec S_i$
is now the spin-$S$ quantum-mechanical operator.
In the special case of $J_1=J_2$, the model can be regarded
as the triangular-ladder model with the nearest-neighbor
antiferromagnetic coupling as considered in \S 2.1.
Here, we extend the triangular-ladder model, or the $J_1=J_2$ model on a
single chain, by
introducing the independent nearest- and next-nearest-neighbor
interactions on a single chain. This
enables us to have a free parameter $j\equiv J_2/J_1$ in the Hamiltonian,
which controls the extent of frustration, and eventually, drives the chiral
order in the ground state. Remember we focus in this section on the
ground-state properties of the model,
since we are interested  in pure quantum
phase transition. The local chirality may be defined as a
quantum-mechanical operator
defined by
\begin{equation}
\kappa_i = S^x_iS^y_{i+1}-S^y_iS^x_{i+1}.
\end{equation}
We note that the possible chiral order of this model has also been studied
analytically on the basis of the field-theoretical method
by Nersesyan {\it et al\/}\cite{Ner}, by Lecheminant
{\it et al\/}\cite{Leche},
and by Kolezhuk\cite{Kole}.

Below, we summarize the properties
revealed mainly  by the exact-diagonalization and DMRG methods
\cite{KKH,HKK1,HKK2}.

\medskip\noindent
i) The case of  $S=1/2$:
With increasing $j$ from $j=0$ to $j=\infty$,
there appear three distinct phases, {\it i.e.\/}, the spin-fluid
phase, the dimer phase and
the gapless chiral phase. The gapless chiral phase
possesses a true chiral LRO with algebraically-decaying spin correlations.
In the numerical accuracy of Ref.\cite{HKK2},
the possible gapped chiral phase (or the chiral dimer
phase) was not identified. The two transitions
associated
with the spin fluid-dimer transition
and with the dimer-gapless chiral transition
are both continuous.
Even for higher half-integer $S\geq 3/2$, qualitative
features of the phase structure
remain the same.

\medskip\noindent
ii) The case of $S=1$: With increasing $j$ from $j=0$ to $j=\infty$,
there appear three distinct phases, {\it i.e.\/}, the Haldane phase,
the gapped chiral phase (or the chiral Haldane phase)
and the gapless chiral phase. The latter two
phases are chiral ordered phases possessing a true chiral LRO.
The gapped chiral phase has exponentially-decaying spin correlations,
while the gapless chiral phase has algebraically-decaying spin correlations.
The two transitions associated
with the Haldane-gapped chiral transition
and with the gapped chiral-gapless chiral transition are both continuous.
We note that the phase structure observed here is quite similar to the one
observed in the classical 2D frustrated {\it XY\/} model as reviewed in
\S 2.2 with varying temperature: Large-, intermediate- and small-$j$ phases
of the quantum 1D model correspond to low-, intermediate- and
high-temperature phases of the classical 2D model. For higher integer
$S\geq 2$, there
appears in addition the spin-fluid phase for smaller values of
$j$. 

\medskip
We note that the results of field-theoretical analyses
and numerical calculations
made so far are consistent with each other on most of the above points.
One ambiguity still being left
might be 
whether there exists a gapped chiral phase (chiral
dimer phase)  for half-odd-integer $S$. Field theory claims that
there should exist such a phase in a narrow interval of $j$ between the
dimer and the gapless chiral phases\cite{Leche},
while the numerical DMRG calculation could
not identify such a phase within the numerical accuracy\cite{HKK2}.
This points needs further clarification.
Quantum chiral order was now studied
for more general types of anisotropy,
{\it e.g.\/}, an {\it XXZ}-type ansiotropy\cite{HKK1,HKK2}
and a single-ion-type anisotropy\cite{Hiki}.
Anyway, owing to the recent analytical and numerical studies,
the existence of quantum chiral order in the frustrated 1D quantum
{\it XY\/}-spin chain has now been well established.
We note in passing that a similar 1D quantum-2D classical analogy
was also examined in terms of a Josephson-junction array in a magnetic
field\cite{Granato,Nishi}.
In this case, the charging effect of superconducting grains plays the role
of the quantum effect.

\bigskip
\noindent
{\bf 4. Concluding remark}
\medskip

A brief review has been given on the recent works on chiral order
in regularly frustrated {\it XY\/} systems both in classical and
quantum cases. In the classical case, chiral transitions of
the  1D, 2D and 3D triangular-lattice antiferromagnets
have been examined. In 1D and 2D, the spin-chirality decoupling phenomenon
takes place at
the transition, while in 3D the chiral transition satisfies
the standard one-length-scaling hypothesis
without the spin-chirality decoupling phenomenon,
though the associated fixed point might well be  novel
because of the underlying
chiral symmetry (a chiral fixed point). In 2D,
there appears a pure chiral phase in an intermediate range of temperature
where the chirality exhibits a LRO with keeping the {\it XY\/} spin
disordered.
In the quantum case, the $T=0$ chiral transition of the 1D spin-$S$
{\it XY\/}
model ($J_1-J_2$ or zigzag chain) has been examined. There exist
two types of chiral phases, gapped and gapless chiral phases, where
the chirality exhibits a LRO. The gapless chiral phase
with algebraically-decaying spin correlations
exists for general
$S$, while the gapped chiral phase
with exponentially-decaying spin correlations
is identified only for integer-$S$.
Analogy between the quantum chiral order in
$D$-dimensions and the thermal chiral order in $D+1$-dimensions
is discussed.

In view of the 1D-quantum and 2D-classical analogy, it would be
natural to extend it to the systems one-dimension higher, {\it i.e.\/}, to
the 2D-quantum and 3D-classical analogy. Then, one expects that
the frustrated 2D quantum system may well exhibit a magnetic phase transition
of chiral universality class.
In fact, such a possibility has already
been examined in several 2D quantum phase
transitions, including those of
Josephson-junction array in a magnetic field\cite{GraKos},
frustrated Heisenberg antiferromagnet\cite{SinghHuse},
square-lattice bilayer Heisenberg model\cite{Matsushita}, and
triangular-lattice bilayer Heisenberg model\cite{SinghElstner}.

Although we have confined ourselves in this review to the chiral order
associated with the
vector chirality in
{\it XY\/}-like systems, there also exist several examples of
Heisenberg-like systems where the
chiral order associated with the scalar chirality takes place.
We have also confined ourselves  to the
regularly frustrated
systems here, neglecting the effects of quenched randomness.
There are several occasions, however,
where the quenched randomness plays an essential
role in the chiral ordering. Examples are the ordering of vector spin glasses, 
including both Heisenberg\cite{KawaSG1,KawaSG2,HukuKawa,KawaImaSG}
and {\it XY\/}\cite{Grempel,KawaLi} spin glasses, 
and of ceramic high-$T_c$ superconductors\cite{KawaLi2,LiNordKawa}.
In fact, quenched randomness generally tends to enhance fluctuations
and serves preferably to cause the
spin-chirality decoupling and  to realize the pure chiral phase
(chiral-glass phase).
The related works have been briefly reviewed in Ref.\cite{Kawareview2}.

Thus, the chiral order, both thermal and quantum,
is likely to be realized in a rather wide class of frustrated systems,
giving rise to intriguing ordering behaviors. Much needs to be done
in the future to fully explore this rich field.

\bigskip
The author is thankful to Dr. M. Kaburagi and Dr. T. Hikihara for
collaboration in the work presented in \S 3.

\bigskip\bigskip

\fussy

\end{document}